\documentclass[12pt,preprint]{aastex}
\usepackage{psfig}

\shorttitle{Cosmology with SNe~II-P}
\shortauthors{Nugent {\it et al.}}

\begin{document}

\newcommand{\bq}{\begin{equation}}
\newcommand{\eq}{\end{equation}}

\bibliographystyle{apj}


\title{Towards a Cosmological Hubble Diagram for Type II-P Supernovae}


\author{Peter Nugent\altaffilmark{1}, Mark Sullivan\altaffilmark{2},
  Richard Ellis\altaffilmark{3}, Avishay Gal-Yam\altaffilmark{3,4},
  Douglas C. Leonard\altaffilmark{3,5}, D. Andrew
  Howell\altaffilmark{2}, Pierre Astier\altaffilmark{6}, Raymond G.
  Carlberg\altaffilmark{2}, Alex Conley\altaffilmark{2}, Sebastien
  Fabbro\altaffilmark{7}, Dominique Fouchez\altaffilmark{8},
  James~D.~Neill\altaffilmark{9}, Reynald Pain\altaffilmark{6}, Kathy
  Perrett\altaffilmark{2}, Chris~J.~Pritchet\altaffilmark{9} and
  Nicolas Regnault\altaffilmark{6}}

\altaffiltext{1}{Lawrence Berkeley National Laboratory, 1 Cyclotron Road,
Berkeley, CA 94720, USA }
\altaffiltext{2}{University of Toronto, 60 St. George Street, Toronto,
ON M5S 3H8, Canada} 
\altaffiltext{3}{California Institute of Technology, E. California
Blvd, Pasadena CA 91125, USA } 
\altaffiltext{4}{Hubble Postdoctoral Fellow}
\altaffiltext{5}{NSF Astronomy and Astrophysics Postdoctoral Fellow}
\altaffiltext{6}{LPNHE, CNRS-IN2P3 and University of Paris VI \& VII,
  75005 Paris, France} 
\altaffiltext{7}{CENTRA - Centro Multidisciplinar de
  Astrof\'{\i}sica, IST, Avenida Rovisco Pais, 1049 Lisbon, Portugal} 
\altaffiltext{8}{CPPM, CNRS-IN2P3 and University Aix Marseille II,
  Case 907, 13288 Marseille Cedex 9, France} 
\altaffiltext{9}{Department of Physics and Astronomy, University of
  Victoria, PO Box 3055, Victoria, BC V8W 3P6, Canada} 

\email{penugent@LBL.gov}


\begin{abstract}
  
  We present the first high-redshift Hubble diagram for Type II-P
  supernovae (SNe\,II-P) based upon five events at redshift up to
  $z\sim 0.3$. This diagram was constructed using photometry from the
  Canada-France-Hawaii Telescope Supernova Legacy Survey and
  absorption line spectroscopy from the Keck observatory. The method
  used to measure distances to these supernovae is based on recent
  work by \citet{hamuysn2p1} and exploits a correlation between the
  absolute brightness of SNe\,II-P and the expansion velocities
  derived from the minimum of the \ion{Fe}{2} 5169\AA\ P-Cygni feature
  observed during the plateau phases.  We present three refinements to
  this method which significantly improve the practicality of
  measuring the distances of SNe\,II-P at cosmologically interesting
  redshifts. These are an extinction correction measurement based on
  the $V-I$ colors at day 50, a cross-correlation measurement for the
  expansion velocity and the ability to extrapolate such velocities
  accurately over almost the entire plateau phase. We apply this
  revised method to our dataset of high-redshift SNe\,II-P and find
  that the resulting Hubble diagram has a scatter of only 0.26
  magnitudes, thus demonstrating the feasibility of measuring the
  expansion history, with present facilities, using a method
  independent of that based upon supernovae of Type Ia.

\end{abstract}


\section{Introduction}

The discovery of a cosmic acceleration based on the analysis of the
Hubble diagram of Type Ia supernovae
\citep[SNe\,Ia;][]{riess_scoop98,42SNe_98} has far-reaching
implications for our understanding of the Universe. While indirect
evidence for the acceleration can be deduced from a combination of
studies of the cosmic microwave background and large scale structure
\citep{2002MNRAS.330L..29E, 2003ApJS..148....1B, 2005ApJ...619..178E}
distance measurements to supernovae provide a valuable direct and
model independent tracer of the evolution of the expansion scale
factor necessary to constrain the nature of the proposed dark energy.
The mystery of dark energy lies at the crossroads of astronomy and
fundamental physics: the former is tasked with measuring its
properties and the latter with explaining its origin.

Systematic uncertainties (rather than statistical errors) may soon
limit SN\,Ia measurements of the expansion rate at $z\sim0.5$
\citep[see][for recent analyses]{knop03,astier05}. A largely
unexplored source of potential bias is evolution in the progenitor
properties and/or the SN explosion. While several programs are
underway to measure, test, and constrain SN\,Ia systematics
\citep{sullivan04,ellis06}, it is highly desirable to consider
independent tests of the cosmology where both the underlying physics
and susceptibility to bias and evolution are different.

As cosmological probes, SNe\,II have lagged behind their brighter and
better calibrated cousins, SNe\,Ia, but their potential has improved
significantly as a result of several recent studies.
\citet{baron_99em,sn93w03,mitchel02}, and two doctoral theses
\citep{mariothesis,dougthesis}, have utilized new samples of SNe\,II
and demonstrated that a subset, the plateau SNe\,II-P, are
particularly promising as distance indicators. From an astrophysical
standpoint, SNe\,II-P hold three advantages over SNe\,Ia as
cosmological probes: (1) Their progenitor stars are well understood,
\citep{2003ApJ...591..288H, 2005PASP..117..121L, 2003MNRAS.343..735S},
(2) the physics of their atmospheres, dominated by hydrogen, is much
simpler to understand and model \citep{baron_99em}, and (3) while
fainter, they are more abundant per unit volume
\citep{2005A&A...433..807M,2005A&A...430...83C}. The two main
disadvantages are that they are on average 1.2 magnitudes fainter in
the optical than SNe\,Ia and that all distance measurements currently
based on SNe\,II-P require a reasonable-quality spectrum of the event.

Unlike the other members of the core-collapse supernova family,
SNe\,II-P maintain a massive hydrogen envelope prior to explosion.
From analyses of their optical light curves and spectra
\citep[e.g.,][]{1994cmls.conf..148C}, they evidently suffer little
subsequent interaction with the surrounding medium -- they are the
result of the putative red supergiant exploding into a near-vacuum.
Recent results from spectropolarimetric studies also suggest that, at
least during the plateau epoch, the ejecta and electron-scattering
photosphere are quite spherical (see \citet{ccpol} and references
therein).

For SNe\,II-P, distance measurements from the spectral expanding
atmosphere method \citep[SEAM; see][]{baron_99em}, the descendent of
the traditional expanding photosphere method \citep[EPM;
see][]{kkepm,bpsphd}, can be replaced by a more practical empirical
method which requires less input data, as developed in
\citet{hamuysn2p1, hamuysn2p2} (hereafter HP02 and H03 respectively).
The approach advocated by HP02 is a particularly significant
development since it is motivated by sound physical principles. In
more luminous supernovae, the hydrogen recombination front is
maintained at higher velocities, pushing the photosphere farther out
in radius.  When a SN\,II-P is on the plateau phase, a period which
lasts for around 100 days, a strong correlation is expected and
observed between the velocity of the weak \ion{Fe}{2} lines near
5000\,\AA\ (which suitably track the electron-scattering photosphere)
and the luminosity.

Calibration of these standardized candles requires line of sight
extinction corrections, which can be determined in one of two ways.
When the supernova leaves the plateau phase, the photospheric
temperature should be close to that of recombination, and thus the
observed color at this point will provide a reddening estimate.
Alternatively, for well-studied SNe\,II-P, one can employ detailed
modeling of a series of high signal-to-noise spectra to determine the
extinction. Here, deviations between the line profiles of selected
absorption features and those expected given the observed spectral
energy distribution (SED) can be used to estimate selective
extinction.  Application of these techniques to 24 SNe\,II-P in the
Hubble flow (Fig.~1 of H03) yields a $I$-band Hubble diagram with a
scatter of only 0.29 magnitudes, corresponding to a precision of 15\%
in distance.

The present paper is concerned with extending the HP02 method so that
it can be used at cosmological distances. Although economical compared
to the SEAM/EPM methods in terms of input data, the HP02 method is
still poorly suited as a basis for verifying the cosmic acceleration
due to the difficult demands on spectroscopy and the photometric
measurements necessary for an extinction correction at high-redshift.
Our discussion is timely because of the feasibility of locating
distant SNe\,II-P from comprehensive `rolling searches' such as the
Canada-France-Hawaii Telescope SN Legacy Survey
\citep[SNLS;][]{astier05}. Such surveys can generate SNe\,II-P with
$0.1 < z < 0.4$ continuously with tight constraints on their explosion
dates and excellent photometric coverage, making pre-scheduled
spectroscopic campaigns to measure the expansion rate during the
plateau phase a practical proposition.

A plan of the paper follows.  In $\S$2 we introduce an enhanced local
dataset suitable for exploring the potential of an improved method and
detail the difficulties we need to overcome to use the HP02 method at
intermediate redshifts. We then introduce two extensions to the method
and show, via our dataset, that we can retain the precision with
indicators of velocity and extinction more suited to high redshift
datasets. In $\S$3, we present the first observations of high-redshift
SNe\,II-P obtained via SNLS and construct a Hubble diagram from these
supernovae which demonstrate the feasibility of locating and studying
high redshift SNe\,II-P for verifying the cosmic acceleration.  We
summarize our conclusions in $\S$4.

\section{Improving the HP02 Method}

The drawbacks of the method introduced by HP02 for studies of
cosmologically distant SNe\,II-P are two-fold.

Foremost, the extinction correction cannot be measured using colors
determined at the end of the plateau phase since its precise timing
would require continuous monitoring, which is impractical for faint
supernovae, and at intermediate redshift, SNe\,II-P discovered prior
to the time they explode will typically be too close to the sun at
this stage in their evolution.  Moreover, corrections for extinction
based on this method in H03 often produced negative results. Likewise,
the signal/noise of plateau spectra will generally not be adequate for
detailed spectrum synthesis modeling of various line profiles in order
to determine the intrinsic SED. We thus seek a more appropriate way to
estimate the line of sight extinction.

Secondly, it is a challenge at moderate redshift to secure accurate
measures of the weak diagnostic \ion{Fe}{2} 5169\AA\ line used to
measure the photospheric expansion during the plateau phase, even with
the largest ground-based telescopes. In particular, as diagnostic
lines in this region of the spectrum are redshifted into the OH forest
beyond $z\simeq$0.5, so it becomes advantageous to explore
alternatives should the \ion{Fe}{2} 5169\AA\ line profile be polluted
by a sky feature. We therefore wish to explore the practicality of
alternative, stronger lines to measure the photospheric expansion.

To explore the possibilities, we have constructed a sample of nearby
SNe\,II-P which includes all of those in H03 for which there is both
$V$- and $I$-band data. We also include SN~2004dh, a SN discovered by
LOSS \citep{sn04dh} and followed as part of the Caltech Core-Collapse
Program \citep[CCCP; see][]{2004AAS...205.4006G,avishay06}, which
leads to a total of 19 SNe\,II-P. We have updated the distances to two
supernovae presented in H03. In the case of SN~1999em we use the {\it
  HST}-determined Cepheid distance to the host \citep{ceph99em}. For
SN~1999gi we use the improved distance derived via a variety of
methods as discussed in \citet{dist99gi}.  These changes have very
little effect on the overall fit as both galaxies are nearby with
large uncertainties in their relative distances. We have also removed
all of the extinction corrections applied in H03.

We now turn to two improvements necessary to enable us to use
SNe\,II-P as distance indicators at moderate redshift.

\subsection{Incorporating $V - I$ Colors}

The first modification of the HP02 method we have explored makes use
of the restframe $V-I$ color during the plateau phase at day 50 to
perform an extinction correction rather than relying on colors at the
end of the plateau phase or on detailed modeling, as in H02. Our goal
is to explore empirically the extent to which we can retain a tight
\ion{Fe}{2} velocity-luminosity relation using the observed $V-I$
color at this epoch.

If we fit for the velocity and luminosity using plateau-phase data,
interpolated to day 50, we find:

\begin{eqnarray}
  M_I = -\alpha\, log_{10}\,(V_{Fe II}/5000) - 1.36 [(V-I) - (V-I)_0] +
  M_{I_0}\label{eq1} 
\end{eqnarray}

\begin{figure}[p]
\psfig{file=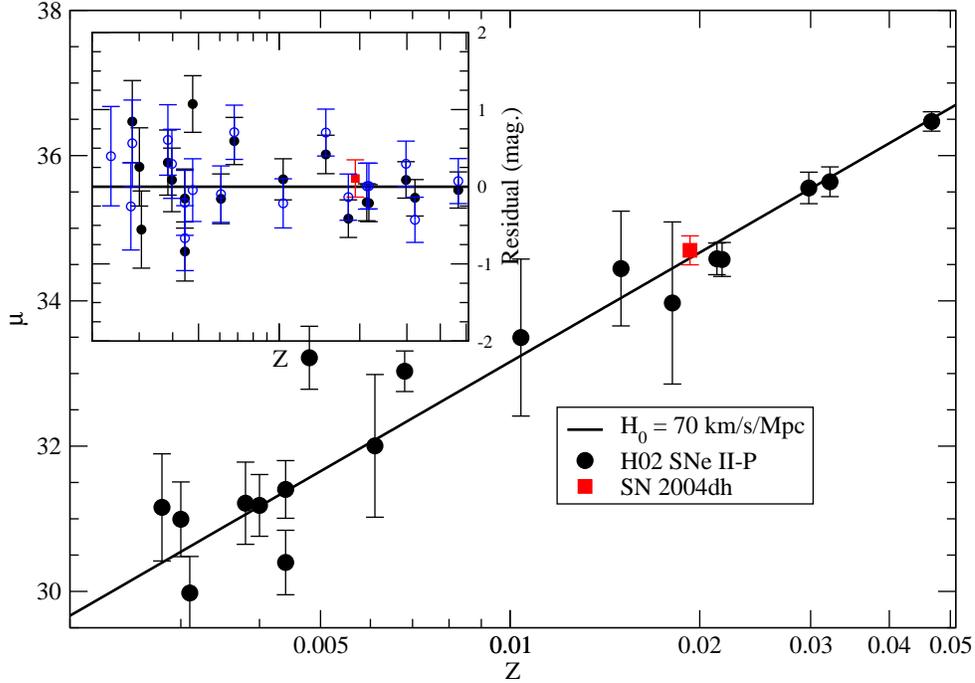,width=6.0in,angle=270}
\caption{A revised Hubble diagram (scaled to $H_0$ = 70 km/s/Mpc) for
  local SNe\,II-P (diamonds and square) using an improved estimator
  for extinction derived from $V-I$ colors and \ion{Fe}{2} velocities
  observed during the plateau phase at day 50. {\it Inset:} The
  residual Hubble diagram for both the revised method (black) and the
  data under the assumption that they are pure standard candles in
  $I$-band, corrected only for MW extinction (blue). The scatter for
  those SN\,II-P in the Hubble flow is reduced from 0.59 to 0.28
  magnitudes, similar to the 0.29 magnitudes originally achieved in
  H03 and based on observables substantially better suited to
  high-redshift observations. The scatter for all SN\,II-P is reduced
  from 1.11 to 0.52 magnitudes.
  \label{lowzhub}}
\end{figure}

Here, as in H03, we have adopted the relative SBF distance scale
\citep{2000ApJ...530..625T}.  The resultant fit to the data yields
$\alpha = 6.69 \pm 0.50$ and $M_{I_0} = -17.49 \pm 0.08$ ($H_0 = 70$
kms sec$^{-1}$ Mpc$^{-1}$) for a $(V-I)_0=0.53$. In this fit we have
employed the standard relationship between the $V-I$ colors for a dust
law with $R_V$ = 3.1 for a SN\,II-P at day 50 ($A_I = 1.36*E_{V-I}$).
As is done in the SN\,Ia studies \citep{guysalt} for the color
stretch-relationship, we have adopted a ridge-line, unextinguished
$(V-I)_0$ color for SN\,II-P of 0.53 magnitudes. Under the assumption
that the extinction laws are similar from galaxy to galaxy, the exact
choice is irrelevant since this term and the $M_{I_0}$ term are
degenerate for the purpose of using these objects as cosmological
probes.  Using this technique we can produce a Hubble diagram in
restframe $I$-band with a scatter of only 0.28 magnitudes
(Figure~\ref{lowzhub}) for those SNe\,II-P in the Hubble flow ($cz$
$>$ 3000 km/s), similar to that found in H03.  The scatter for all
SN\,II-P is reduced from 1.11 to 0.52 magnitudes. Crucially, we find
there is no advantage in using reddening estimators based on late-time
color measures or on detailed modeling of the spectroscopic data.

To address the robustness of this simple approach, we investigated
solutions where we permit an additional parameter based upon a
possible relationship between the \ion{Fe}{2} velocity and the
dereddened $V-I$ color. One could easily envision a correlation that
brighter SN\,II-P were faster {\it and} bluer. However, the
correlation we find is weak and lacks any statistical significance.

We note that for $\alpha \ne 5.0$ ($\alpha = 5$ implies that the
luminosity follows the square of the radius only), there presumably
does exist an additional correlation between the effective temperature
and density of the expanding photosphere and its radius, so
conceivably there is some scope for further improvement in defining
Eq.~\ref{eq1}. Future work, such as the ongoing research by the CCCP
and the Carnegie Supernova Project (CSP) \citep{2005ASPC..339...50F},
will be helpful in clarifying the possibilities and allow us to
address the strongly correlated issues of extinction corrections and
velocity-dependent colors.

\subsection{Using Alternative Lines to Diagnose the Photospheric Expansion}
\label{fe2} 

\begin{figure}[p]
\psfig{file=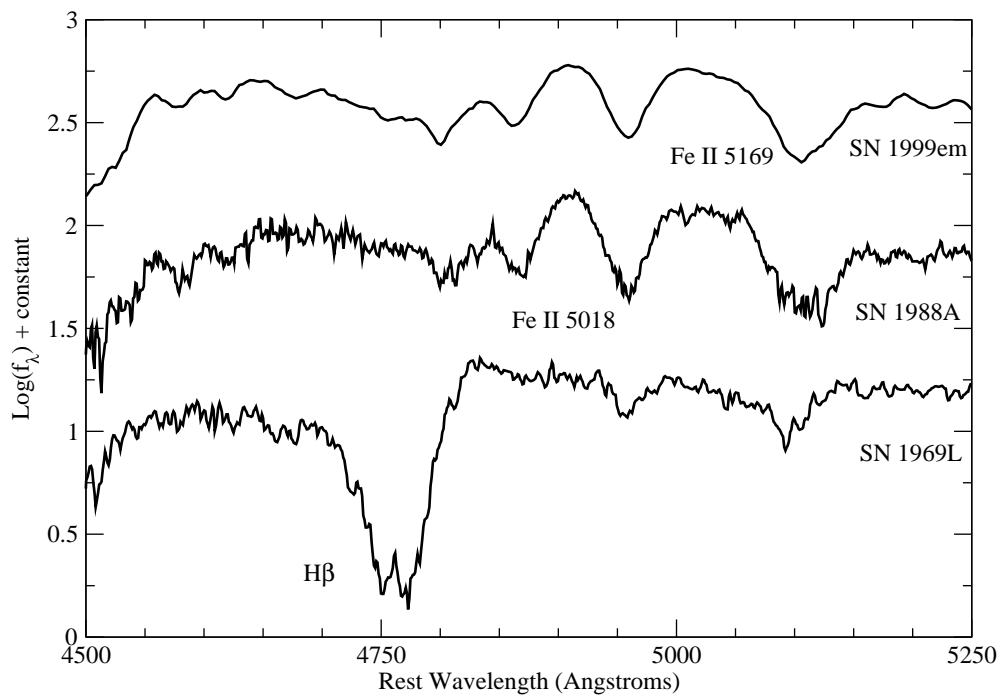,width=6.0in,angle=270}
\caption{A comparison of two \ion{Fe}{2}-dominated spectra to an
  $H\beta$-dominated spectrum. This demarcation is defined by the
  equivalent width of $H\beta$ being greater than the sum of the
  \ion{Fe}{2} features. Depending on the supernova and its phase, any
  one of these features could potentially be the strongest and/or the
  most easily measured (due to sky lines or host galaxy contamination)
  and thus could be used at high redshift.\label{sn2p_abs}}
\end{figure}

We now explore whether alternative absorption lines can be used in
addition to \ion{Fe}{2} 5169\AA\ as diagnostics of the photospheric
expansion velocity.  Over the wavelength range 4500-5500\AA\ there
exist several \ion{Fe}{2} lines (the strongest of which are at 4924,
5018 and 5169\AA). Moreover, often the $H\beta$~4861\AA\ absorption
line is prominent.  Depending on the supernova and its phase, the
relative strengths of these lines can vary considerably (see
Figure~\ref{sn2p_abs}). In general, at earlier epochs $H\beta$ tends
to dominate while a later epochs the \ion{Fe}{2} lines are the
dominant absorption features over this wavelength range.  Accordingly,
if at higher redshift we are able to compare to any or all of these
lines we can minimize the effects of the interference with OH night
sky features and any underlying host galaxy contamination.

In order to explore possible systematic effects arising from the use
of using different absorption lines as measures of the expansion, we
will first examine trends in local data. To do this we assembled a
library of SNe\,II-P spectra based on data from the SUSPECT
database\footnote{see: {\rm
    http://bruford.nhn.ou.edu/\~{}suspect/index1.html}} for:
SNe\,1969L, 1988A, 1988H, 1993W, 1999gi and
1999em\citep{1991sos..conf..339B,
  1993MNRAS.265..471T,dist99gi,ceph99em,hamuy_99em,sn93w03} as well as
data from the first year of the CCCP: SNe~2004A, 2004T, 2004dh,
2004du, 2004em and 2004et \citep{cccp_first}. A particularly important
question is whether it is safe to use $H\beta$ as a diagnostic rather
than the \ion{Fe}{2} lines, given the former is often quite strong.
In what follows, we have divided the local dataset into two
categories: those dominated by the $H\beta$ feature and those
dominated by the \ion{Fe}{2} features (defined according to whether
the equivalent width of $H\beta$ is respectively greater than or less
than that of the sum of the \ion{Fe}{2} features). The motivation here
is two-fold. Firstly, unlike $H\beta$, in the case of the
\ion{Fe}{2}-dominated spectra, the weak lines form at the same
location in the atmosphere near the electron scattering photosphere.
Thus we might expect any (or all) of them to be acceptable for
measuring the velocity.  Secondly, as we will later test a novel
cross-correlation method across this wavelength range, it is very
helpful to understand possible systematic velocity differences between
H$\beta$ and the \ion{Fe}{2} lines. For both \ion{Fe}{2} 5169\AA\ and
H$\beta$, we measured absorption velocities using a routine in which
we first subtract the continuum, take the wavelength derivative and
then fit for the wavelength where this derivative changes sign.

\begin{figure}[p]
\psfig{file=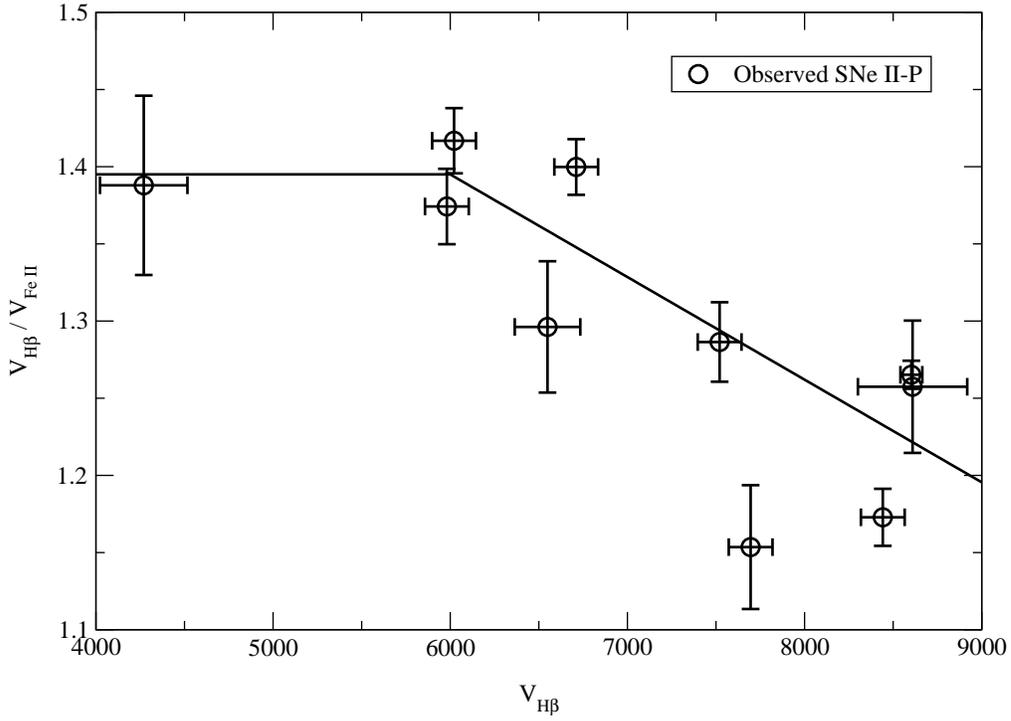,width=6.0in,angle=270}
\caption{The ratio of the velocities of $H\beta$ to \ion{Fe}{2}
  5169\AA\ vs. the $H\beta$ velocity in the $H\beta$-dominated
  spectra. We fit this ratio in two pieces, a constant for velocities
  below 6000~km/s ($ratio=1.395$) and linearly for higher velocities
  ($ratio = 1.395 - 6.489*10^{-5}(V(H\beta) - 6000)$). The linear
  portion is the only one relevant to our analysis as our
  high-redshift data lie within these velocities. The dispersion
  about this relationship ($\pm$0.054) translates to an uncertainty in
  measuring the \ion{Fe}{2} velocity from $H\beta$ of $\sim$300 km/s.
  See text for a full discussion and explanation of this trend.
  \label{velrat}}
\end{figure}

Figure~\ref{velrat} shows that, in the $H\beta$-dominated spectra, the
velocity of $H\beta$ is significantly higher than that of \ion{Fe}{2}
5169\AA\ with the trend that the ratio of $H\beta$ to \ion{Fe}{2}
drops towards unity only at very high velocities. Clearly simply
replacing \ion{Fe}{2} absorption velocities with measures of H$\beta$
would lead to erroneous results.  However, if we fit this ratio in two
sections, a constant for velocities below 6000~km/s ($ratio=1.395$)
and a linear decline for higher velocities ($ratio = 1.395 -
6.489*10^{-5}(V(H\beta) - 6000)$), the dispersion about this
relationship ($\pm$0.054) translates to an uncertainty in estimating
the \ion{Fe}{2} velocity from $H\beta$ measures of only $\sim$300
km/s.

Of course, before accepting such an empirical correction, it is
important to understand the trend physically. To accomplish this, we
examined spectrum synthesis fits to model SNe\,II-P discussed by
\citet{baron_99em}.  The predicted trend is qualitatively similar in
form to that observed and, in particular, reproduces the decline in
the ratio observed at higher velocities.  The model spectra with
higher effective temperatures were in general from earlier epochs with
higher velocities, while the cooler ones were later with lower
velocities.  The optical depth of $H\beta$ is strongly tied to the
level of ionization of hydrogen. Thus as the temperature drops, the
velocities fall, less hydrogen is ionized and the optical depth of
$H\beta$ increases significantly compared to the weak \ion{Fe}{2}
lines. This in turn increases the ratio.

In order to maximize the information content in either the series of
weak \ion{Fe}{2} lines or in the $H\beta$-dominated spectra, we then
examined a cross-correlation analysis across the wavelength range
4500-5500\AA\ . In the \ion{Fe}{2} dominated spectra we simply
compared the measured \ion{Fe}{2} 5169\AA\ velocities to those
obtained in the cross-correlation analysis. For the $H\beta$ dominated
spectra we compare the measured $H\beta$ velocities and then reduced
these to equivalent \ion{Fe}{2} 5169\AA\ velocities using the fit to
the ratio defined above. In performing cross-correlations, all spectra
are continuum subtracted and analyzed in log-wavelength space in the
manner described by \citet{1979AJ.....84.1511T}.

\begin{figure}[p]
\psfig{file=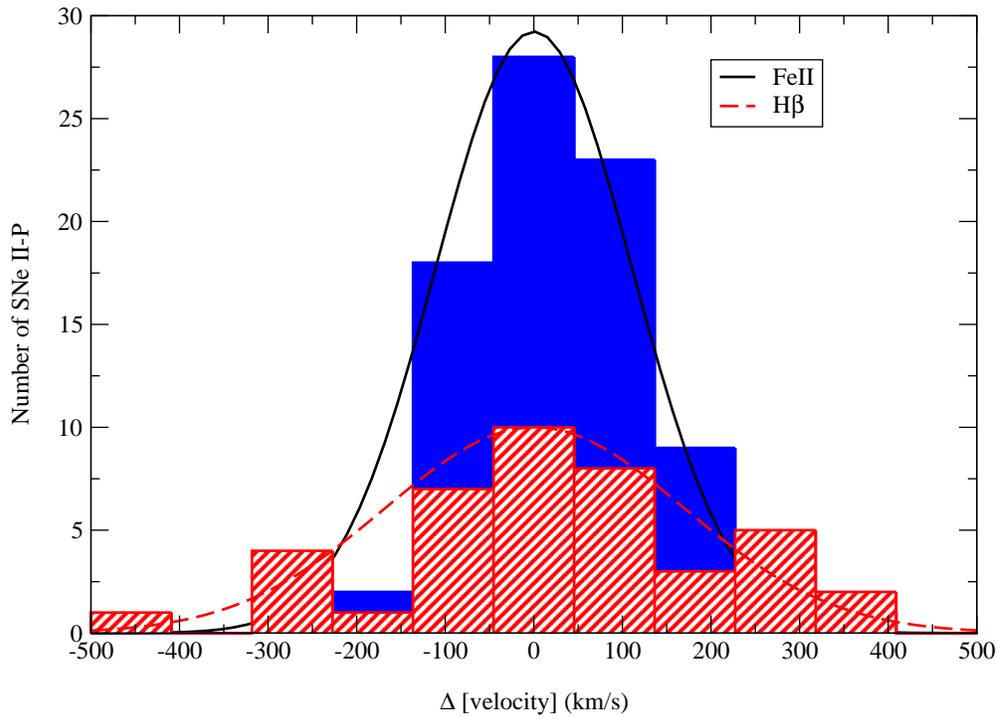,width=6.0in,angle=270}
\caption{A histogram of the difference in measured velocity vs. the
  velocity derived from the cross-correlation technique for
  \ion{Fe}{2} 5169\AA\ and $H\beta$. Here we have compared each
  SN\,II-P in our library to all the others, splitting the comparison
  library for SN\,II-P to those that are dominated by $H\beta$ and
  those by \ion{Fe}{2}.  The measured dispersion is 161~km/s for the
  $H\beta$ subset and 108~km/s for the \ion{Fe}{2} subset.
  \label{cc_comp}}
\end{figure}

A measure of the potential bias and systematic uncertainty arising in
the cross-correlation method can be obtained by considering the
distribution of velocity differences obtained by multiple template
comparisons within our local dataset (Figure~\ref{cc_comp}). Each
local supernova, when compared to the average of the cross-correlation
calculated velocities, shows differences of typically $<$150~km/s.
This is a measure of {\it template mismatch} for the local sample in
the wavelength region sample, i.e. in individual SNe\,II-P, the
strongest features present can form at slightly different velocities
than \ion{Fe}{2} 5169\AA\ in addition to measurement uncertainties for
the minimum of that line in our spectral library. This bias is
recorded for each of our template spectra and subtracted off when used
in the cross-correlation measurement. The resultant measured
dispersion in comparing one of our template spectra to the rest in
their respective sets is 160~km/s for $H\beta$-dominated spectra and
110~km/s for the \ion{Fe}{2}-dominated spectra

Of course these tests calibrate systematic uncertainties that are
dominant only for low redshift SNe\,IIP. At high redshift we can
expect an additional component from the lower signal to noise implicit
in the fainter sources. To estimate how effective this method will be
at intermediate redshift, where our signal-to-noise is typically
between 8-20 per 2\AA\ over the relevant wavelength range, we
performed a Monte Carlo equivalent of the above test by degrading the
signal-to-noise of the nearby spectra to 7 per 2\AA. The resulting
dispersion for individual measurements increased by only 15~km
s$^{-1}$, clearly showing that the cross-correlation method is
sufficiently accurate for our purpose.

\begin{figure}[p]
\psfig{file=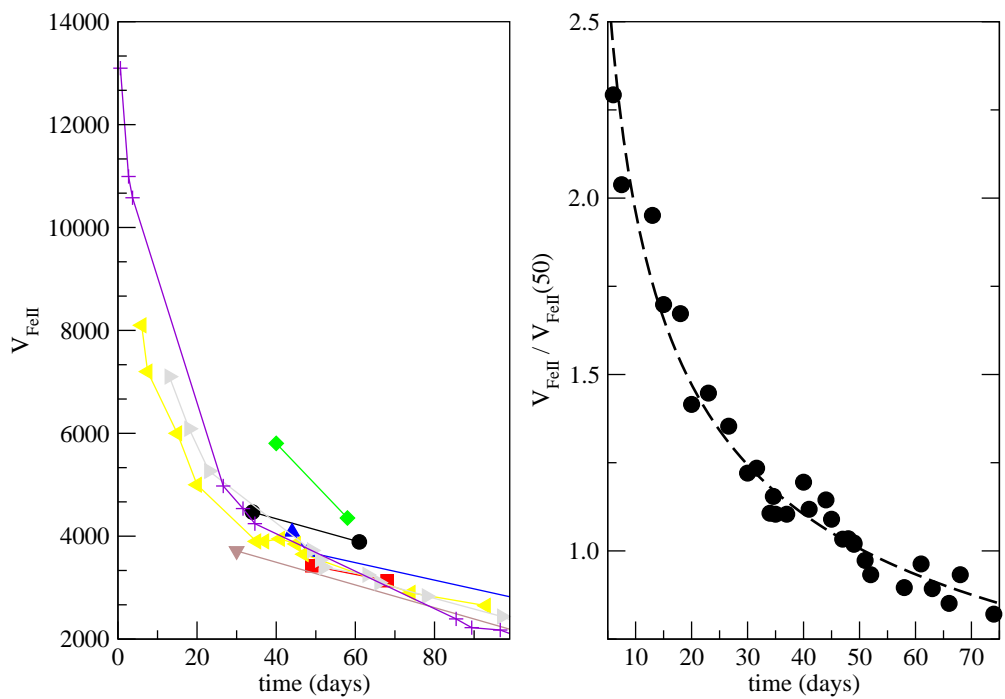,width=6.0in,angle=270}
\caption{{\it Left:} A plot of the velocity of \ion{Fe}{2} 5169\AA\
  vs.~time for the published SNe\,II-P (each symbol represents an
  individual supernova). {\it Right:} A plot of the individual
  SNe\,II-P with their velocities normalized to day 50, along with the
  best fit for the evolution of the decline covering the epochs for
  our high-redshift supernovae (day 9--75). The fit is a power law of
  the form $V(50) = V(t)*(t/50)^{0.464 \pm 0.017}$.\label{fe2pwr}}
\end{figure}

The final problem we must overcome in facilitating a cosmological
application of SNe\,II-P concerns the fact that it is impractical to
schedule, in advance, a spectroscopic night at day 50 for each
supernova of interest. Thus we need to explore how well we can
extrapolate the velocity from a given epoch to day 50. While published
data with the run of velocity with time is limited, that available to
us indicates that the time dependence of the \ion{Fe}{2} 5169\AA\ 
velocity can be well represented by a power law of the form:
\begin{eqnarray}
V(50) = V(t)*(t/50)^{0.464 \pm 0.017}
\label{eq2}
\end{eqnarray}
(see Figure~\ref{fe2pwr}).  Extrapolation for epochs between 9 and 75
days to day 50 can be made quite reliably adding an uncertainty of
$<$175 km/s. Adding the various uncertainties in quadrature indicates
expansion velocities of distant SNe\,II-P should be typically accurate
to $\simeq$220 km/s.

\section{A Cosmological Hubble Diagram Based on a Sample of High
Redshift SNe\,II-P} 

In the previous section, we have shown that extinction corrections
using $V-I$ colors during the plateau phase at day 50 and concurrent
expansion velocities determined using a variety of prominent
absorption lines can be used to generalize and extend the important
distance determination method initially proposed by HP02.  As
SNe\,II-P are now being found to redshifts $z\simeq 0.4$ in rolling
searches such as the SNLS \citep{sullivan04}, we can now apply our
method to these distant supernovae to independently verify the cosmic
acceleration.

\begin{figure}[p]
\psfig{file=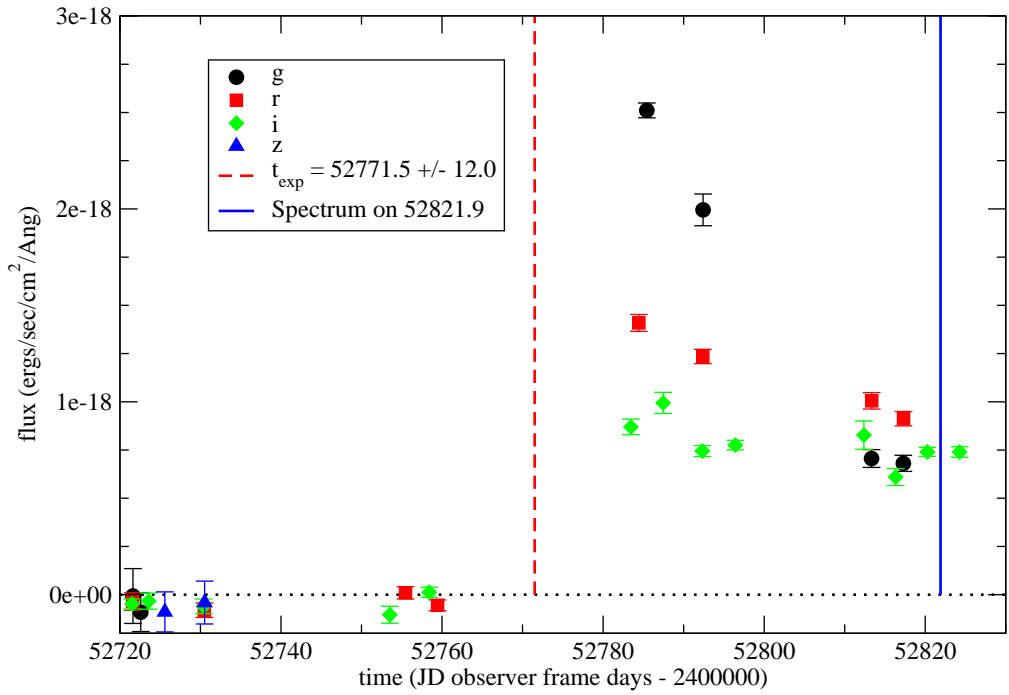,width=6.0in,angle=270}
\caption{\label{03d3ce} The SNLS $g'r'i'z'$ lightcurves (in
  $f_{\lambda}$) for SNLS-03D3ce at a $z=0.2881$. Unfortunately this SN
  was not observed in $z'$ on the plateau. At this redshift $z'$
  overlaps nicely with restframe $I$-band, thus the $V-I$ color had to
  be extrapolated from bluer colors than desired for this SN\,II-P.}
\end{figure}

\begin{figure}[p]
\psfig{file=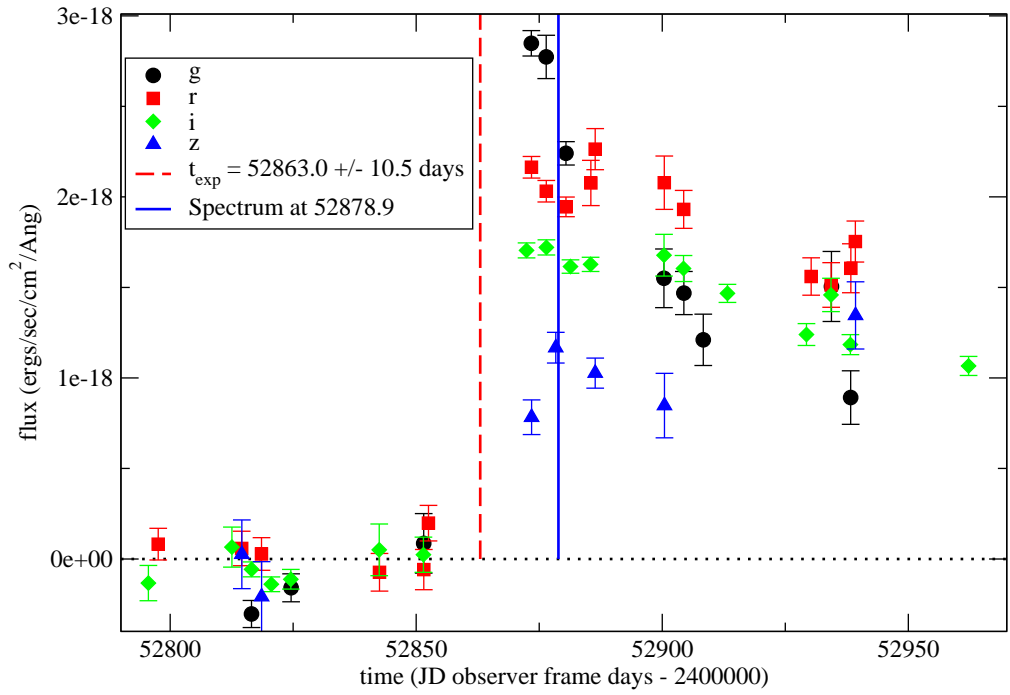,width=6.0in,angle=270}
\caption{\label{03d4cw}The SNLS $g'r'i'z'$ lightcurves (in
  $f_{\lambda}$) for SNLS-03D4cw at a $z=0.1543$.} 
\end{figure}

\begin{figure}[p]
\psfig{file=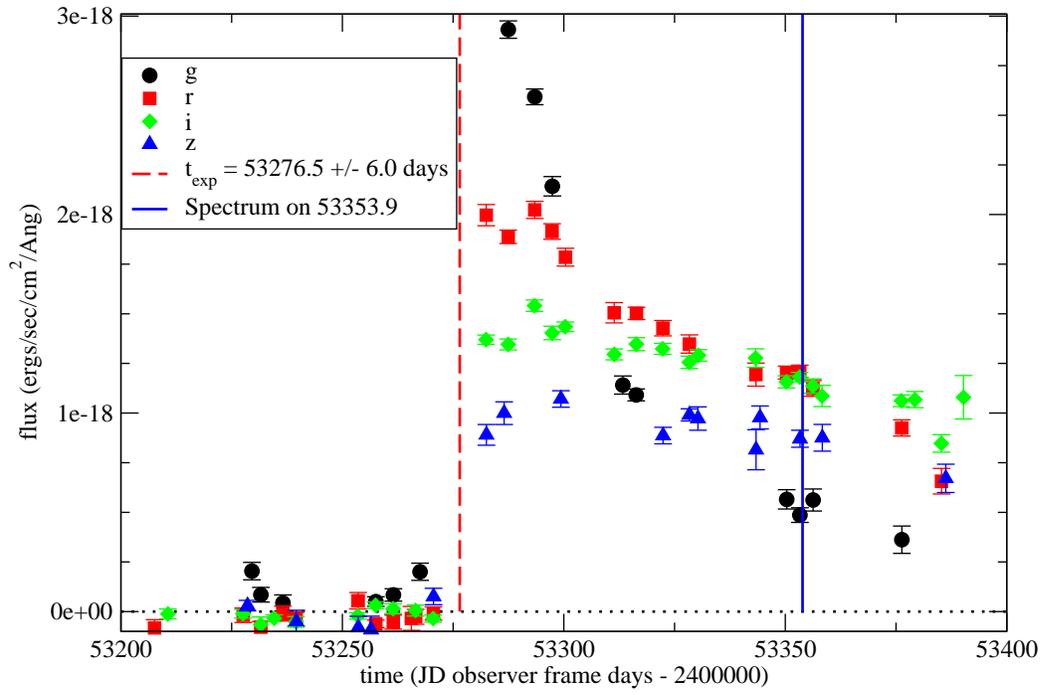,width=6.0in,angle=270}
\caption{\label{04d1ln}The SNLS $g'r'i'z'$ lightcurves (in
  $f_{\lambda}$) for SNLS-04D1ln at a $z=0.2078$.} 
\end{figure}

\begin{figure}[p]
\psfig{file=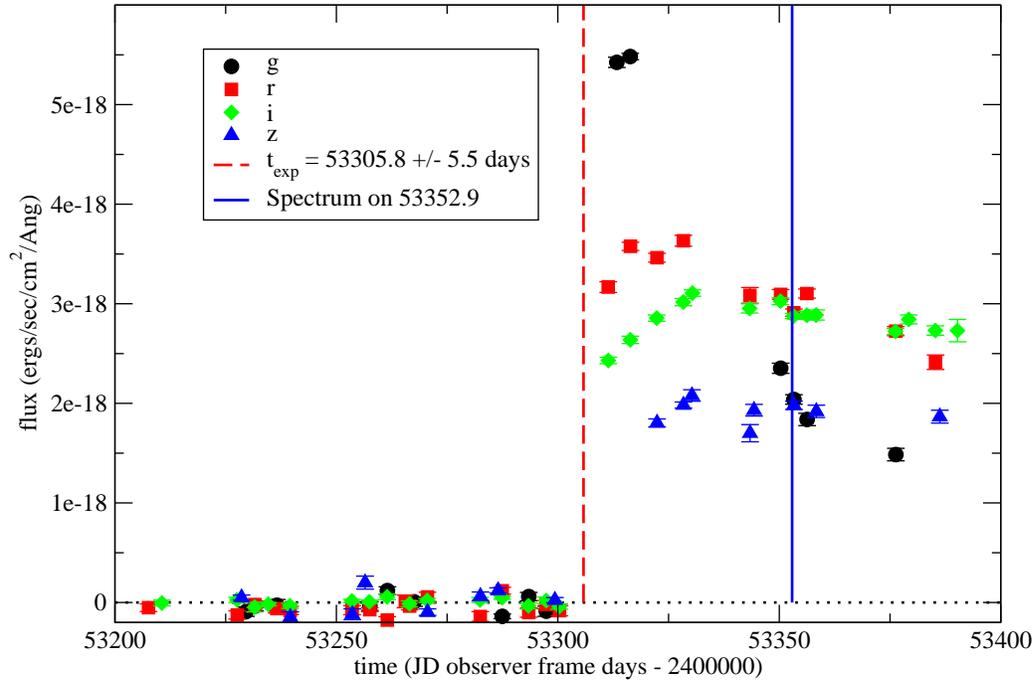,width=6.0in,angle=270}
\caption{\label{04d1pj}The SNLS $g'r'i'z'$ lightcurves (in
  $f_{\lambda}$) for SNLS-04D1pj at a $z=0.1556$.} 
\end{figure}

Starting in 2003, we began using the Low Resolution Imaging
Spectrometer [LRIS; \citet{1995PASP..107..375O}], a double-arm
spectrograph mounted on the 10-m Keck~I telescope, to observe
SNLS-discovered supernovae. Although our primary program initially
targeted SNe\,Ia \citep{ellis06}, during the first two years we also
successfully studied five moderate-redshift ($0.1 < z < 0.3$)
SNe\,II-P, identified from the SNLS rolling search via their light
curves. These events were typically observed five times per lunation
in $g'r'i'z'$ \citep{astier05} with coverage starting before explosion
and extending throughout the plateau phase (see
Figures~\ref{03d3ce}--\ref{04d4fu}).

We measure $g'r'i'z'$ fluxes of the SNLS supernovae using a
PSF-fitting technique. The first step is to perform a photometric
alignment of each supernova observation to secondary standards within
each field using a multiplicative scaling factor, and assign an
astrometric solution using custom-built astrometric reference
catalogs. Deep "references" are constructed to contain no supernova
light, and are resampled to the pixel co-ordinate system of each
observation. The references are then PSF-matched and subtracted from
each individual observation, leaving a difference image containing
only supernova photons.  The fluxes are then estimated by fitting a
PSF profile measured from stars in the un-subtracted image at the
known position of the supernova, with the weighting in each pixel
equal to $1/\sigma^2$, where $\sigma$ is calculated from the
background sky noise and the Poisson noise from the supernova and host
photons. An average flux measurement on each night of observation is
made using a sigma-clipped weighted average of the individual data
frames.

Keck spectra were taken on the plateau over a range in time of $15 < t
< 75$ restframe days (see Figure~\ref{spectra}). The Keck-LRIS
observing configuration used the 560 dichroic, with the 600l/mm grism
blazed at 4000\AA\ on the blue side, and the 400l/mm grating blazed at
8500\AA\ on the red side. The grating central wavelength was adjusted
to ensure a complete wavelength coverage from 3500\AA\ to 9200\AA.
Depending on seeing conditions, either a 1\arcsec\ or 0.7\arcsec\ slit
was used, and the supernova was dithered between each 1800s exposure
by 3-5\arcsec. Typical total exposure times were 2-3 hours, and all
data were taken at the parallactic angle. All spectroscopic data was
processed using a pipeline developed by one of us (MS), which performs
basic bias subtraction and flat-fielding, as well as a fringe
subtraction on the red side using a fringe frame constructed from an
entire night's data. Each exposure was extracted separately on the red
and blue sides, wavelength and flux-calibrated, corrected for
atmospheric extinction, and optimally combined with the other
exposures to form one continuous spectrum.

To calculate the restframe $V-I$ colors during the plateau phase, we
first made an estimate of the explosion date of the supernova
conservatively based on the midpoint between the last non-detection
and the first detection of the supernova in any filter during the
rolling search. We then interpolated the $g'r'i'z'$ colors at day
$50(1+z)$. These colors were used to ``warp'' a day 50 template
SN\,II-P spectrum, redshifted appropriately, following the protocol
described in \citet{2002PASP..114..803N} with the exception that here
we use spline fits to the underlying colors to adjust the template
rather than a reddening law. This spectrum was then de-redshifted and
the restframe $V-I$ color of the supernova was calculated.

The \ion{Fe}{2} velocities were measured via the cross-correlation
technique introduced in \S~\ref{fe2}, extrapolated to restframe day 50
using Eq.~\ref{eq2}.  In determining the velocity and its uncertainty
from the full range of local templates, we adopted the {\it Median
  Absolute Deviation} method, which provides a particularly robust
measure of the dispersion when there are a few significant outliers.
The distribution of cross-correlation velocities is calculated using
all local templates and the median of the distribution estimated. The
absolute modulus of the deviation from the median velocity is
calculated and that distribution is medianed and 3$\sigma$ clipped. We
then take the mean of the remaining velocities and the rms about this
mean as our velocity and uncertainty.  Typically, this procedure
reduced the comparison library by $<$25\%.  The correlations and
resulting best fits can be seen in Figure~\ref{ccplot} and are listed
in Table~\ref{cctabhb} and~\ref{cctabfe}.

Appropriately dealing with uncertainties in the above measurements,
and evaluating their implications on the inferred distances is an
important aspect of our analysis. An uncertainty in the explosion date
will propagate to one in the distance determination. On the one hand
the expansion velocity will decline (implying a lower luminosity) as
the explosion date is moved forward, whereas the SN will get redder
with time on the plateau, implying more extinction (and a more
luminous event). To quantify such biases, we devised a Monte Carlo
simulation incorporating these uncertainties. The results in
Table~\ref{highztab} summarize both the measurements and their
uncertainties as a result of these simulations. To illustrate the
potential value of our Hubble diagram, we performed a cosmological fit
using both these and and the low-redshift supernova
(Table~\ref{lowztab}) under the assumption of the flat cosmological
model found in \citet{astier05}.

The values of the fitting parameters in this joint fit of both the
high and low redshift SNe\,II-P changed to $\alpha=5.81$ and
$M_{I_0}=17.52$. This value of $\alpha$ is lower by 1.5-$\sigma$
compared to the fit for only the low-redshift SNe\,II-P, see
Figure~\ref{confid}. The observed scatter was 0.26 magnitudes for all
supernovae in the Hubble flow. The Hubble diagram is presented in
Figure~\ref{highzhub}. The reduced $\chi^2$ is 1.8 which implies an
intrinsic dispersion of 0.12 magnitudes, though this number is quite
tentative due to the low number statistics involved in this analysis.

Two of the problems facing this dataset of SNe\,II-P are the
data-analysis inconsistencies between the low and high-redshift data
and small number statistics. The low redshift data assembled in H03
measured and then extrapolated/interpolated the velocities in a manner
different than the one presented here (though for high-quality data
with strong time constraints, there should be little to no difference
between the techniques). In addition, a large fraction of the data in
H03 involves SNe\,II-P not in the Hubble flow. Our five additional
high-redshift supernovae, and one low-redshift SN\,II-P, increased
this pool of data from 8 to 14. To address these issues and understand
the nature of our dataset, its limitations and the robustness of this
result we performed a bootstrap resampling (with replacement) study of
the 24 SNe\,II-P. In Figure~\ref{boot} we present the results of this
study for the variable $\alpha$ in Eq.~\ref{eq1} and the resultant
rms. The rms peaks at 0.26 with a dispersion of 0.05 magnitudes and is
well behaved. $\alpha$ peaks at 6.0 with a dispersion about the
central peak of 0.5 (similar to that found in the formal $\chi^2$
analysis), but shows a long tail to higher values potentially
indicating the effect of small number statistics on this analysis.

\section{Discussion}

In this paper we have presented the first Hubble diagram for SNe\,II-P
at cosmologically significant redshifts. The rms scatter of this
method, 13\% in distance, compares favorably to the 7-10\% scatter
seen in the SN\,Ia measurements
\citep{riess_scoop98,42SNe_98,astier05}. This research has reached the
stage of the early SNe\,Ia studies. Further improvements can now be
sought with a view to reducing the scatter and increasing the
cosmological power of the high-redshift data.

\begin{figure}[p]
\psfig{file=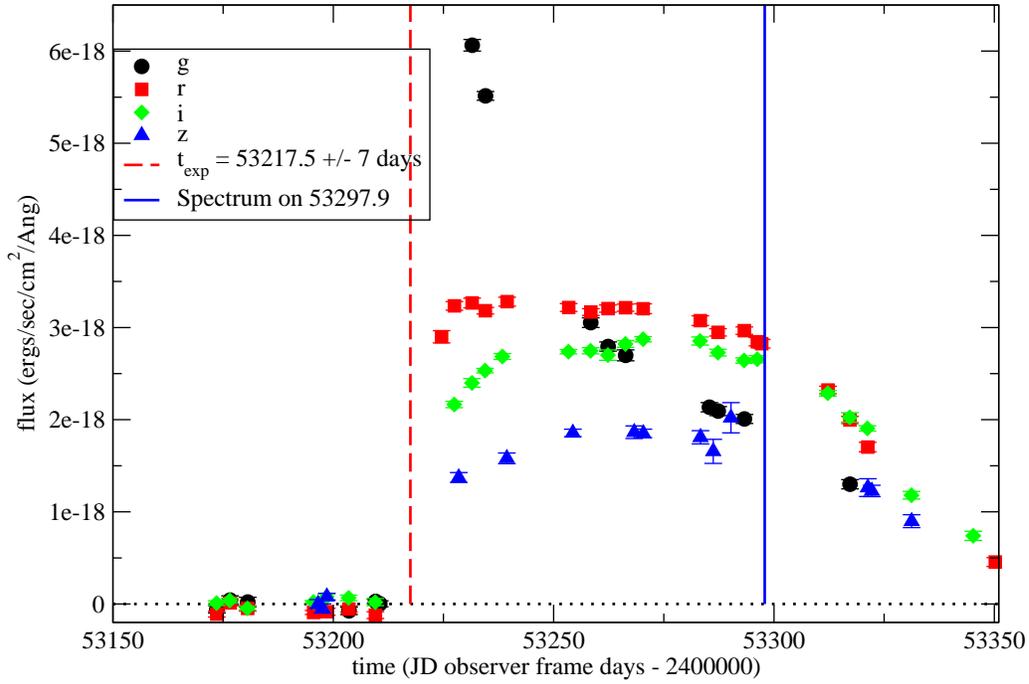,width=6.0in,angle=270}
\caption{\label{04d4fu}The SNLS $g'r'i'z'$ lightcurves (in
  $f_{\lambda}$) for SNLS-04D4fu at a $z=0.1330$. This supernova was
  caught just after shock-breakout as the $g'r'i'z'$ colors
  (uncorrected for extinction) lead to a temperature $>$ than
  20,000~K.}
\end{figure}

\begin{figure}[p]
\psfig{file=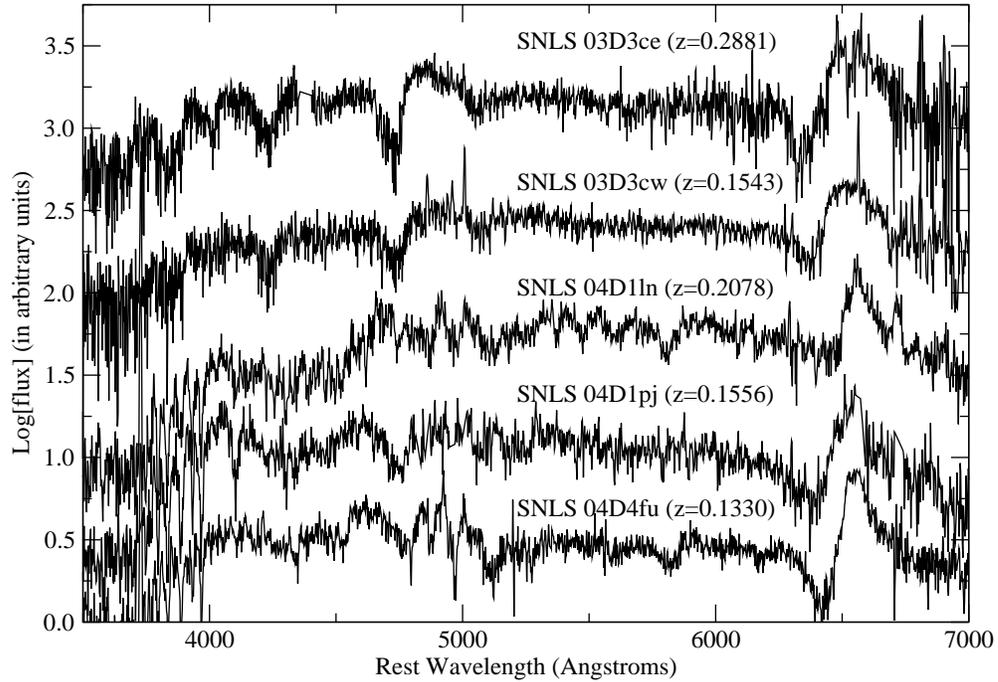,width=6.0in,angle=270}
\caption{\label{spectra} The restframe spectra of the
  high-redshift SNe\,II-P observed at Keck+LRIS in 2003 and 2004. From
  top to bottom they are: SNLS~03D3ce, 03D4cw, 04D1ln, 04D1pj and
  04D4fu. The quality of these spectra is such that the measurement of
  the \ion{Fe}{2} velocities, based on the technique described in
  \S~\ref{fe2} over the range 4500-5500~\AA, is good to $\sim$250
  km/s.}
\end{figure}

\begin{figure}[p]
\psfig{file=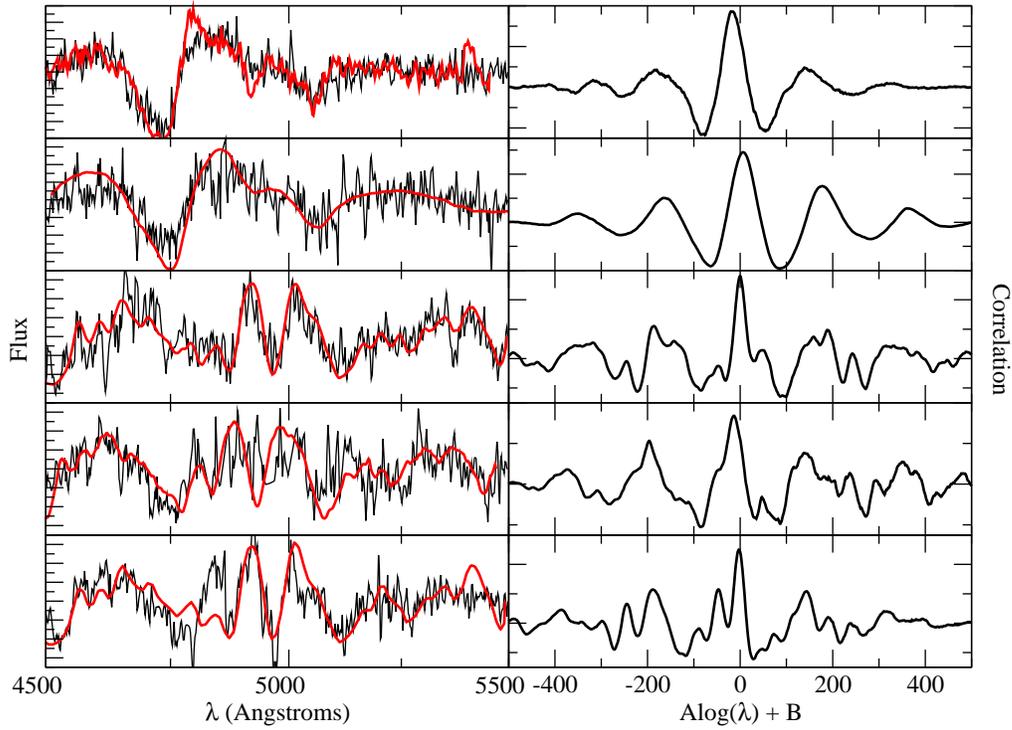,width=6.0in,angle=270}
\caption{\label{ccplot} {\it Left-hand-side}: A plot of the best
  fit template spectrum (thick line) and the SNLS SNe\,II-P (thin
  line) used in the cross-correlation technique. {\it Right-hand-side}:
  The resultant correlation. Plotted from top to bottom are:
  SNLS-03D3ce, 03D4cw, 04D1ln, 04D1pj and 04D4fu.}
\end{figure}

\begin{figure}[p]
\psfig{file=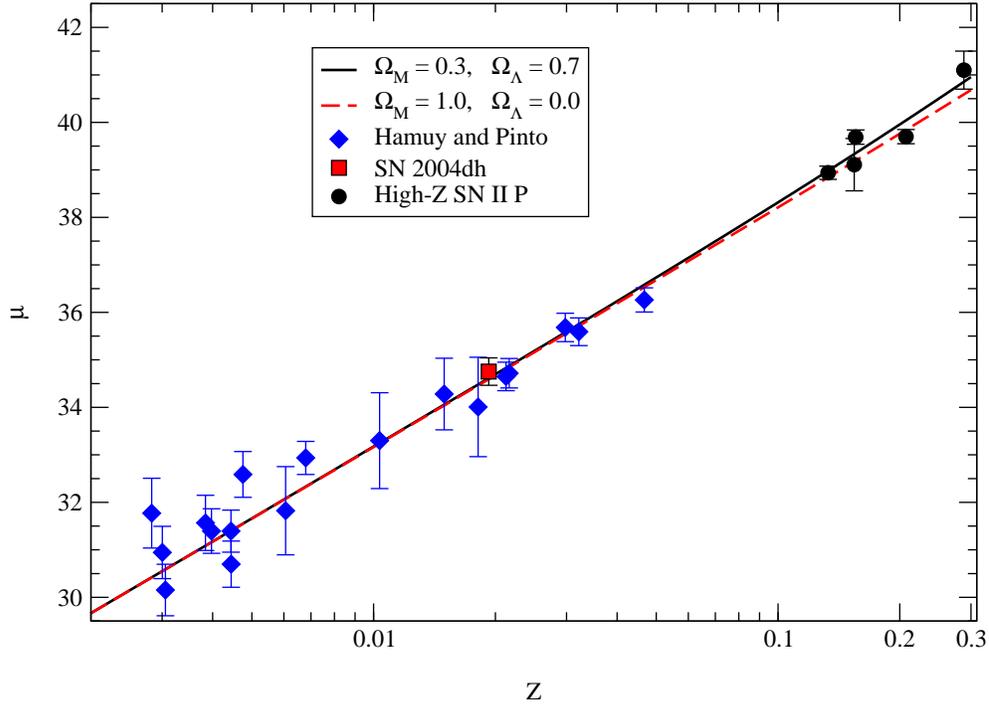,width=6.0in,angle=270}
\caption{The Hubble diagram for both the local SNe\,II-P (diamonds and
  square) and the high-redshift SNLS SNe\,II-P (circles) observed
  spectroscopically with Keck+LRIS. The observed scatter for the
  supernova in the Hubble flow is 0.26 magnitudes with a reduced
  $\chi^2$ of 1.8, which is indicates an intrinsic dispersion of 0.12
  magnitudes. To understand the current power of this technique we
  have over-plotted two differing Hubble lines for a flat cosmology
  with $\Omega_M=1$ and 0.3.
  \label{highzhub}}
\end{figure}

\begin{figure}[p]
\psfig{file=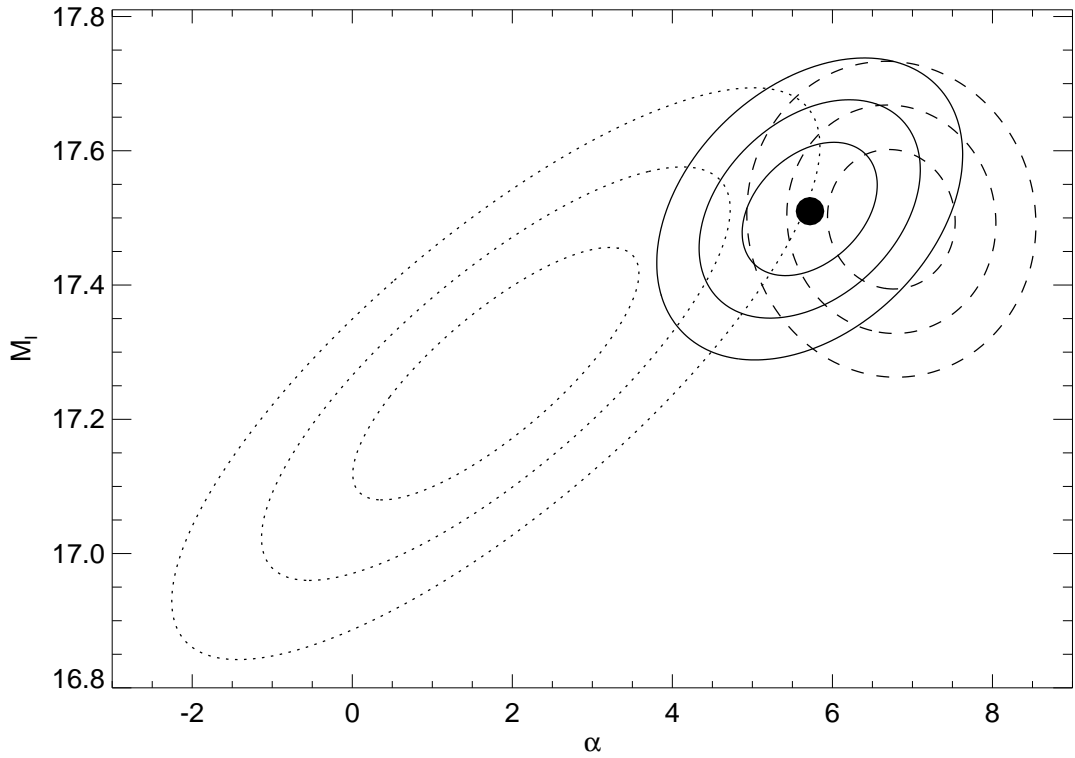,width=6.0in}
\caption{The 1, 2 and 3-$\sigma$ confidence contours in the
  $\alpha$--$M_{I_0}$ plane for the low-redshift only (dashed),
  high-redshift only (dotted) and combined SNe\,II-P data sets
  (solid). Due to low-number statistics at both high and low-redshift
  these parameters are still poorly determined for precision
  cosmology. Future research by the CCCP and CSP will improve our
  understanding of this relationship considerably.
  \label{confid}}
\end{figure}

\begin{figure}[p]
\psfig{file=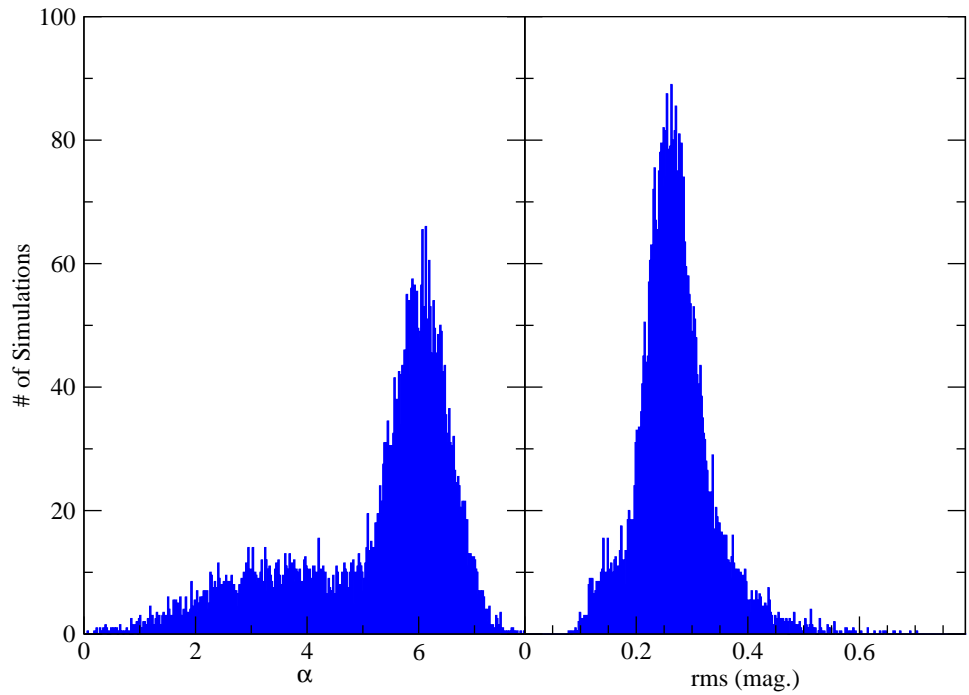,width=6.0in,angle=270}
\caption{A bootstrap resampling (with replacement) study of the
  variable $\alpha$ and the rms of the 24 SNe\,II-P presented in this
  paper. While the rms is nicely behaved, $\alpha$ indicates that our
  dataset may be susceptible to small number statistics.
  \label{boot}}
\end{figure}

Examples of potential improvements include construction of new
lightcurve templates to improve estimates of the explosion date. In
Figure~\ref{04d4fu} the color of the SNLS-04D4fu, upon initial
detection, implies temperatures of $\sim$20,000~K, suggesting this
event was observed shortly after explosion and shock breakout, yet
presently we do not incorporate this information to help constrain the
explosion date.  Furthermore, while we observed only 5 SNe\,II-P
during the first two years of the SNLS program, the current
photometric datasets contain many well observed lightcurves which,
once redshifts have been obtained for the host galaxies, will allow us
to place tight constraints on how various SNe\,II-P evolve as a
function of time and thus better measure both their plateau colors and
the date of explosion.

\begin{figure}[p]
\psfig{file=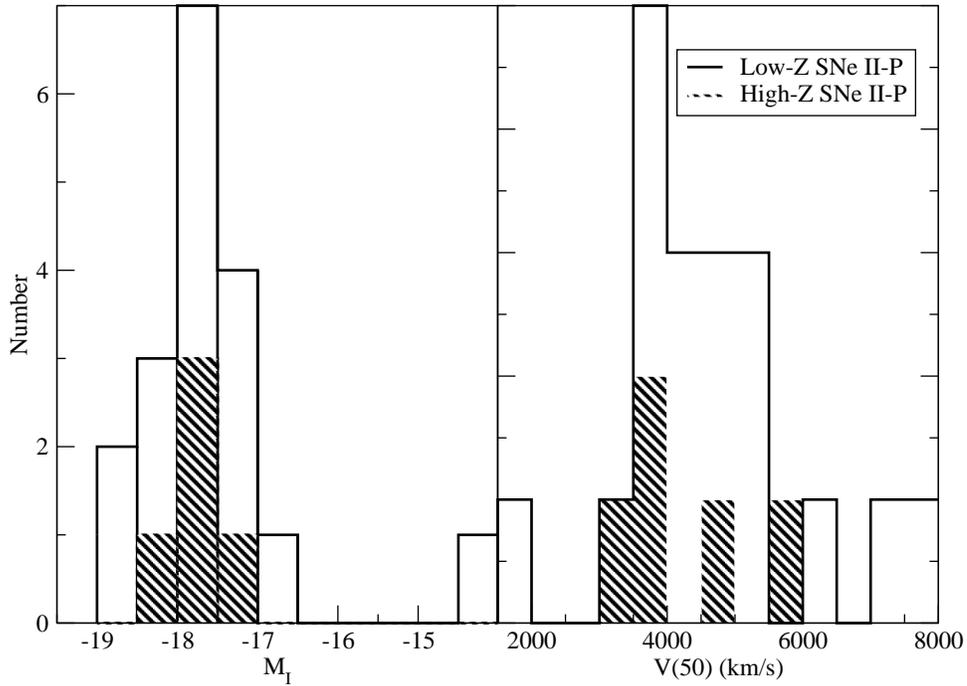,width=6.0in,angle=270}
\caption{Histograms of the observed absolute magnitude in $I$-band,
  uncorrected for extinction, and the day 50 velocities, for both the
  high-redshift (filled) and low-redshift (open) SNe\,II-P.  Future
  datasets will have to address the important issue of Malmquist bias,
  a question beyond the scope of this paper.
  \label{malm}}
\end{figure}

Additionally, dozens more nearby Hubble-flow SNe\,II-P have been
observed by the CCCP and CSP both photometrically and
spectroscopically.  These data sets will allow us to better understand
the velocity evolution of the spectroscopic features on the plateau
and see if it is possible to use different photometric bands to
standardize the SNe\,II-P while we improve our current methods.  One
question that this and future SNe\,II-P datasets must address is that
of Malmquist bias. The two effects this can have on our analysis are a
bias on the parameters we have determined in Eq.~\ref{eq1} and a bias
on the discovery of lower luminosity events given the magnitude
limited search. For the latter bias the important parameters to
measure are the completeness limit of the search and the intrinsic
dispersion on the corrected SNe\,II-P magnitudes (see \citet{42SNe_98}
and references therein). From the data presented in this paper (see
Figure~\ref{malm}) it is not clear whether or not Malmquist bias is a
major issue or even which dataset, the low-redshift or high-redshift
one, suffers from it more. Both the distributions in absolute
magnitude and the velocity at day 50 are similar for each set.

Given multi-color lightcurves ($g'r'i'z'$) provided by SNLS, we
conclude that this method {\it as-is} can be practically applied,
using existing instrumentation, to $z=0.3$. Over this redshift range
we could detect the cosmic acceleration at $>$95\% level, independent
of any other constraints on the cosmological parameters, with
$\simeq$12 additional high-redshift SNe\,II-P coupled with the five
low-redshift CCCP SNe\,II-P observed this past year. Extending these
measurements to $z=0.5$ for the SNLS supernovae would require $J$-band
imaging with {\it HST/NICMOS} for average luminosity SNe\,II-P.
However, due to the wide dispersion in their apparent luminosity, a
large fraction of this distribution could be observed in the infrared
with 8-10 meter ground-based telescopes.

Exploring the utility of measuring distances to SNe\,II-P has
potential benefits well beyond simply verifying, independently, the
acceleration seen at redshifts $z<$1. Several plausible models for the
time evolution of the dark energy require distance measures to
$z\simeq$2 and beyond \citep{2003PhRvD..67h1303L}. At such high
redshifts, other cosmological probes may become less effective than at
$z\le1$. Weak lensing, for example, will suffer from the loss of
suitable lenses, and, while evidence at $z<1$ suggests some fraction
of SNe\,Ia explode with very short delay-times and hence will be
abundant at high-redshift \citep{2005A&A...433..807M,sullivan06}, the
efficiency and metallicity dependence of the SN\,Ia progenitor system
is still in doubt and may curtail their production
\citep[e.g.][]{kobayashi_98}. However, current models for the cosmic
star-formation history predict an abundant source of SNe\,II at these
epochs and future facilities, such as the proposed {\it JDEM}
telescope, in concert with {\it JWST} and/or future 30-m telescopes
such as TMT, could potentially use SNe\,II-P to determine distances at
these very high redshifts.

\acknowledgments

Based on observations obtained with MegaPrime/MegaCam, a joint project
of CFHT and CEA/DAPNIA, at the Canada-France-Hawaii Telescope (CFHT)
which is operated by the National Research Council (NRC) of Canada,
the Institut National des Science de l'Univers of the Centre National
de la Recherche Scientifique (CNRS) of France, and the University of
Hawaii. This work is based in part on data products produced at the
Canadian Astronomy Data Centre as part of the Canada-France-Hawaii
Telescope Legacy Survey, a collaborative project of NRC and CNRS.
Some of the data presented herein were obtained at the W.M. Keck
Observatory, which is operated as a scientific partnership among the
California Institute of Technology, the University of California and
the National Aeronautics and Space Administration. The Observatory was
made possible by the generous financial support of the W.M. Keck
Foundation. The authors wish to recognize and acknowledge the very
significant cultural role and reverence that the summit of Mauna Kea
has always had within the indigenous Hawaiian community.  We are most
fortunate to have the opportunity to conduct observations from this
mountain. PEN acknowledges support from NASA ATP and LTSA grants.  RSE
acknowledges financial support from DOE.  AG acknowledges support by
NASA through Hubble Fellowship grant \#HST-HF-01158.01-A awarded by
STScI, which is operated by AURA, Inc., for NASA, under contract NAS
5-26555. DCL is supported by a National Science Foundation (NSF)
Astronomy and Astrophysics Postdoctoral Fellowship under award
AST-0401479. This research used resources of the National Energy
Research Scientific Computing Center, which is supported by the Office
of Science of the U.S. Department of Energy under Contract No.
DE-AC03-76SF00098. We thank them for a generous allocation of
computing time.

\clearpage

\begin{deluxetable}{ccccc}
\tablewidth{0pt}
\tabletypesize{\footnotesize}
\tablecaption{Cross-Correlation Velocity Analysis for $H\beta$
  Dominated Spectra}
\tablehead{\colhead{SNLS Name} & Template Spectrum (Epoch) &
  $H\beta$ Velocity (km/s) &  $H\beta$/\ion{Fe}{2} Ratio &
  \ion{Fe}{2} Velocity (km/s)} 
\startdata
03D3ce & SN 1999em (15) & 8137.6 $\pm$ 42.0 & 1.256 $\pm$ 0.054 &
6479.0 $\pm$ 350\\
03D4cw & SN 2004T  (15) & 7355.4 $\pm$ 56.3 & 1.307 $\pm$ 0.054 &
5627.7 $\pm$ 304\\
\enddata 
\tablecomments{The best match nearby template spectrum from our
  library along with the the resultant median absolute deviation
  velocity measurement, with a 3 median-absolute-deviations cut, for
  the $H\beta$ dominated SNLS supernovae. Also listed are the $H\beta$
  to \ion{Fe}{2} ratios and the resultant \ion{Fe}{2} velocities.
  Approximate epochs for the nearby SNe\,II-P are in
  parens.\label{cctabhb}}
\end{deluxetable}

\clearpage

\begin{deluxetable}{ccc}
\tablewidth{0pt}
\tabletypesize{\footnotesize}
\tablecaption{Cross-Correlation Velocity Analysis for \ion{Fe}{2}
  Dominated Spectra}
\tablehead{\colhead{SNLS Name} & Template Spectrum (Epoch) &
  \ion{Fe}{2} Velocity (km/s)} 
\startdata
04D1ln & SN 1999em (61) & 3201.8 $\pm$ 86.9 \\
04D1pj & SN 2004em (35) & 5478.5 $\pm$ 72.5 \\
04D4fu & SN 2004A  (50) & 3280.5 $\pm$ 35.1 \\
\enddata 
\tablecomments{The best match nearby template spectrum from our
  library along with the the resultant median absolute deviation
  velocity measurement, with a 3 median-absolute-deviations cut, for
  the \ion{Fe}{2} dominated SNLS supernovae. Approximate epochs for the
  nearby SNe\,II-P are in parens.\label{cctabfe}}
\end{deluxetable}

\clearpage

\begin{deluxetable}{lccccccc}
\tablewidth{0pt}
\tabletypesize{\footnotesize}
\tablecaption{Measurements for the SNLS high-redshift SNe\,II-P}
\tablehead{\colhead{SNLS Name} & \colhead{$z$} &\colhead{$t_{expl}$} &
  \colhead{$t_{spec}$}  & \colhead{$V_{Fe~II}$ (km/s)}
  &\colhead{$I$} & \colhead{$V-I$} & \colhead{$\mu$}} 
\startdata
03D3ce & 0.2881 & 2771.5 $\pm$ 12.0 & 2821.9 & 5762 $\pm$ 522 & 
23.45 $\pm$ 0.50 & 0.39 $\pm$ 0.50 & 41.10 $\pm$ 0.40\\
03D4cw & 0.1543 & 2863.0 $\pm$ 10.5 & 2878.9 & 3067 $\pm$ 660 &
22.33 $\pm$ 0.09 & 0.90 $\pm$ 0.10 & 39.11 $\pm$ 0.55\\
04D1ln & 0.2077 & 3276.5 $\pm$  6.0 & 3353.9 & 3593 $\pm$ 159 &
22.79 $\pm$ 0.05 & 0.70 $\pm$ 0.07 & 39.70 $\pm$ 0.15\\
04D1pj & 0.1556 & 3305.8 $\pm$  5.5 & 3352.9 & 4981 $\pm$ 214 &
21.99 $\pm$ 0.04 & 0.66 $\pm$ 0.06 & 39.69 $\pm$ 0.15\\
04D4fu & 0.1330 & 3217.5 $\pm$  7.0 & 3297.9 & 3861 $\pm$ 150 &
21.99 $\pm$ 0.04 & 0.59 $\pm$ 0.05 & 38.94 $\pm$ 0.14\\
\enddata 

\tablecomments{All dates are in the observer-frame and are expressed
  as (JD-2450000) and the uncertainties in the explosion date are not
  corrected for time dilation. The magnitudes have been corrected for
  MW extinction along the line of sight. The large uncertainties for
  SNLS-03D3ce are due to the fact that it was not observed in $z'$,
  which at its redshift overlaps with restframe $I$-band. For this
  supernova we derived it's $V-I$ color by finding the best match in
  our template day 50 spectra to its observed $g'r'i'$ magnitudes.  We
  then set the dispersion in the $V-I$ color to 50\%, which covers the
  range of all uncorrected colors for the SNe\,II-P presented in this
  paper. Our Monte Carlo (see text) was used to derive both the peak
  values and the uncertainties in the distance modulii. In general,
  the major contribution to the uncertainty in the distance are (in
  decreasing significance): explosion date uncertainty, length of
  extrapolation to day 50, velocity uncertainty and photometry.}

\label{highztab}
\end{deluxetable}

\begin{deluxetable}{lccccc}
  \tablewidth{0pt} \tabletypesize{\footnotesize}
  \tablecaption{Measurements for the Local \& Hubble-Flow SNe\,II-P}
  \tablehead{\colhead{IAUC Name} & \colhead{$cz$} &
    \colhead{$V_{Fe~II}$ (km/s)} &\colhead{$I$} & \colhead{$V-I$} &
    \colhead{$\mu$}} \startdata
  1986I  &  1333 & 3623 $\pm $300  & 13.98 $\pm$ 0.09 & 0.45 $\pm$ 0.22 & 30.58 $\pm$ 0.43 \\
  1989L  &  1332 & 3529 $\pm $300  & 14.47 $\pm$ 0.05 & 0.88 $\pm$ 0.07 & 31.60 $\pm$ 0.38 \\
  1990E  &  1426 & 5324 $\pm $300  & 14.51 $\pm$ 0.20 & 1.31 $\pm$ 0.28 & 33.24 $\pm$ 0.43 \\
  1990K  &  1818 & 6142 $\pm $2000 & 13.87 $\pm$ 0.05 & 0.58 $\pm$ 0.21 & 31.96 $\pm$ 0.87 \\
  1991al &  4484 & 7330 $\pm $2000 & 16.06 $\pm$ 0.05 & 0.39 $\pm$ 0.07 & 34.33 $\pm$ 0.70 \\
  1991G  &  1152 & 3347 $\pm $500  & 15.01 $\pm$ 0.09 & 0.45 $\pm$ 0.11 & 31.43 $\pm$ 0.53 \\
  1992af &  5438 & 5322 $\pm $2000 & 16.46 $\pm$ 0.20 & 0.43 $\pm$ 0.28 & 33.99 $\pm$ 0.99 \\
  1992am & 14009 & 7868 $\pm $300  & 17.90 $\pm$ 0.05 & 0.38 $\pm$ 0.07 & 36.33 $\pm$ 0.12 \\
  1992ba &  1192 & 3523 $\pm $300  & 14.65 $\pm$ 0.05 & 0.59 $\pm$ 0.07 & 31.38 $\pm$ 0.41 \\
  1993A  &  8933 & 4290 $\pm $300  & 18.56 $\pm$ 0.05 & 0.51 $\pm$ 0.07 & 35.67 $\pm$ 0.20 \\
  1993S  &  9649 & 4569 $\pm $300  & 18.22 $\pm$ 0.05 & 0.69 $\pm$ 0.07 & 35.73 $\pm$ 0.18 \\
  1999br &   848 & 1545 $\pm $300  & 16.67 $\pm$ 0.05 & 0.84 $\pm$ 0.07 & 31.70 $\pm$ 0.69 \\
  1999ca &  3105 & 5353 $\pm $2000 & 15.56 $\pm$ 0.05 & 0.73 $\pm$ 0.07 & 33.51 $\pm$ 0.95 \\
  1999cr &  6376 & 4389 $\pm $300  & 17.44 $\pm$ 0.05 & 0.56 $\pm$ 0.07 & 34.68 $\pm$ 0.20 \\
  1999eg &  6494 & 4012 $\pm $300  & 17.72 $\pm$ 0.05 & 0.55 $\pm$ 0.07 & 34.71 $\pm$ 0.21 \\
  1999em &   917 & 3757 $\pm $300  & 13.28 $\pm$ 0.05 & 0.57 $\pm$ 0.07 & 30.15 $\pm$ 0.49 \\
  1999gi &   901 & 3617 $\pm $300  & 13.95 $\pm$ 0.05 & 0.91 $\pm$ 0.07 & 31.18 $\pm$ 0.50 \\
  2000cb &  2038 & 4732 $\pm $300  & 15.48 $\pm$ 0.05 & 0.71 $\pm$ 0.07 & 33.10 $\pm$ 0.26 \\
  2004dh &  5778 & 4990 $\pm $300  & 17.32 $\pm$ 0.05 & 0.47 $\pm$ 0.07 & 34.74 $\pm$ 0.18 \\
  \enddata \tablecomments{Data from \citet{hamuysn2p1, hamuysn2p2}        
    with modifications to SNe~1999em and 1999gi as stated in the text.    
    SN~2004dh is from preliminary work by the CCCP                        
    \citep{2004AAS...205.4006G,avishay06}. All magnitudes have been       
    corrected for MW extinction along the line of sight.}                 
\label{lowztab}
\end{deluxetable}

\clearpage

\clearpage

\clearpage

\clearpage

\clearpage

\clearpage

\clearpage

\clearpage

\clearpage

\clearpage

\clearpage

\clearpage

\clearpage

\clearpage

\clearpage

\clearpage

\end{document}